\documentclass{article}

\usepackage{arxiv}

\usepackage[title]{appendix}
\usepackage[utf8]{inputenc} 
\usepackage[T1]{fontenc}    
\usepackage[numbers]{natbib}
\usepackage{booktabs}       
\usepackage{amsfonts}       
\usepackage{amsmath,bm,upgreek}
\usepackage{hyperref}       
\usepackage{url}            
\usepackage{nicefrac}       
\usepackage{microtype}      
\usepackage{cleveref}       
\usepackage{graphicx}
\usepackage{doi}
\usepackage{subfig}

\graphicspath{ {./images/}}

\DeclareMathOperator{\T}{\mathsf{T}}
\newcommand{\dd}[1]{\mathrm{d}#1}
\newcommand{\Delb}{\Delta \mathbf{b}}
\newcommand{\pxb}{p(\mathbf{x|b})}

\newcommand{\py}{p(\mathbf{y})}
\newcommand{\pyb}{p(\mathbf{y|b})}
\newcommand{\pybb}{p(\mathbf{y|b}+\Delta\mathbf{b})}
\newcommand{\Err}{\mathbb E \left[ \mathbf{rr}^{\T} \right]}

\title{A general framework for probabilistic sensitivity analysis with respect to distribution parameters}

\date{February 1, 2023}

\author{ \href{https://orcid.org/0000-0001-8323-7406}{\includegraphics[scale=0.01]{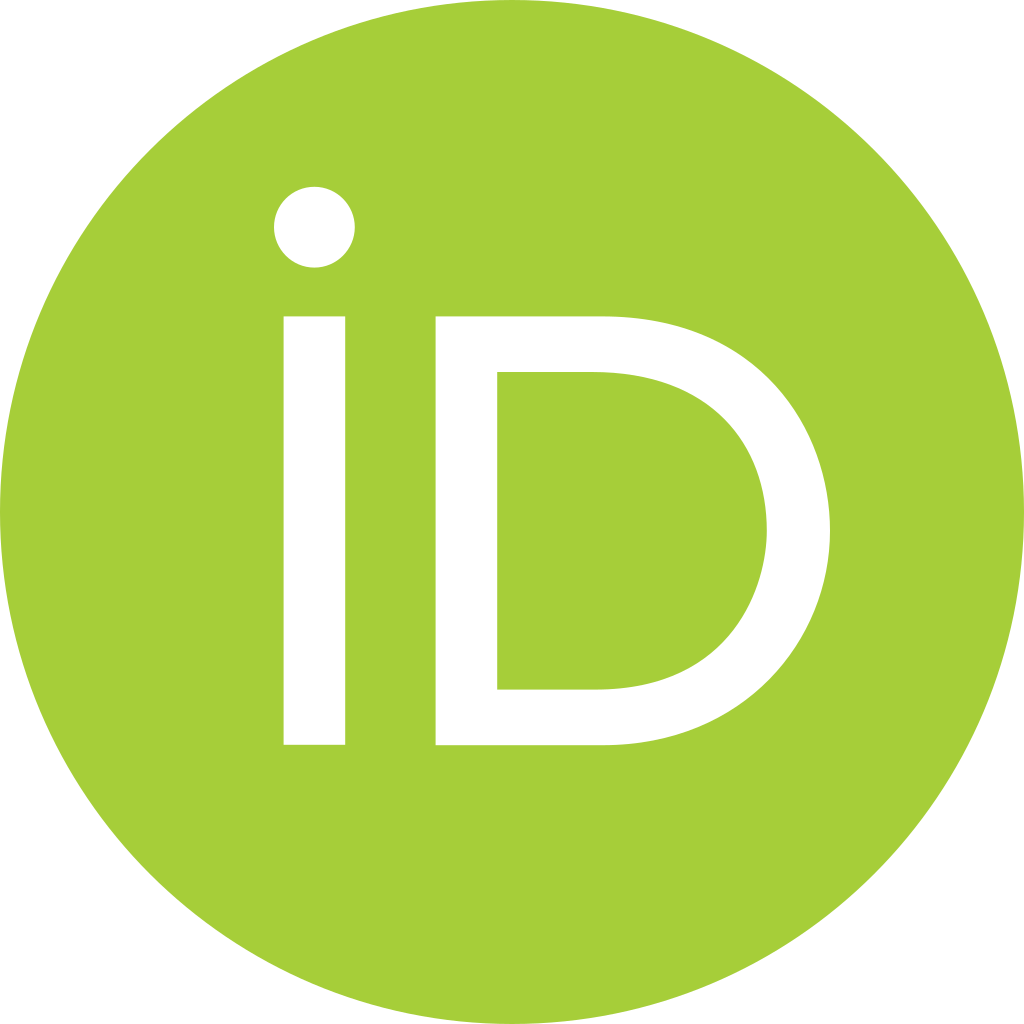}\hspace{1mm}Jiannan Yang}
    \\
  	Department of Engineering\\
	University of Cambridge\\
	Trumpington Street, Cambridge CB2 1PZ, UK\\
	\href{mailto:jy419@cam.ac.uk}{jy419@cam.ac.uk}\\
}

\renewcommand{\shorttitle}{\textit{Probabilistic Sensitivity Framework}}

\hypersetup{
pdftitle={},
pdfsubject={},
pdfauthor={Jiannan Yang},
pdfkeywords={},
}

\begin{document}
\maketitle

\begin{abstract}
Probabilistic sensitivity analysis identifies the influential uncertain input to guide decision-making. We propose a general sensitivity framework with respect to the input distribution parameters that unifies a wide range of sensitivity measures, including information theoretical metrics such as the Fisher information. The framework is derived analytically via a constrained maximization and the sensitivity analysis is reformulated into an eigenvalue problem. There are only two main steps to implement the sensitivity framework utilising the likelihood ratio/score function method, a Monte Carlo type sampling followed by solving an eigenvalue equation. The resulting eigenvectors then provide the directions for simultaneous variations of the input parameters and guide the focus to perturb uncertainty the most. Not only is it conceptually simple, but numerical examples demonstrate that the proposed framework also provides new sensitivity insights, such as the combined sensitivity of multiple correlated uncertainty metrics, robust sensitivity analysis with an entropic constraint, and approximation of deterministic sensitivities. Three different examples, ranging from a simple cantilever beam to an offshore marine riser, are used to demonstrate the potential applications of the proposed sensitivity framework to applied mechanics problems.

\end{abstract}



\section{Introduction}
\label{sec:1}
The use of mathematical models to simulate real-world phenomena is firmly established in many areas of science and technology. The input data for the models are often uncertain, as they could be from multiple sources and of different levels of relevance. The uncertain inputs of a mathematical model induce uncertainties in the output and sensitivity analysis identifies the influential inputs to guide decision-making. A broad range of approaches can be found in the literature, but in practice, the input uncertainties are commonly quantified by a joint probability distribution. The analysis of the input and output relationship in this probabilistic setting is called probabilistic sensitivity analysis \citep{oakley_probabilistic_2004}. 

A suitable measure can be used to summarise the induced output uncertainties. Commonly used metrics are the (central) moment functions of the uncertain output, such as the mean and variance, and the probability of failure, i.e., the probability that the random output would exceed a certain threshold. In addition, the average uncertainty or information content can be measured using entropy that is based on the entire distribution function of the random output \cite{cover_elements_2006}. The probabilistic sensitivity analysis then examines the relationship between the uncertain input and the induced uncertainty of the output. In particular, we are interested in identifying which input parameters would impact the output metrics the most, i.e., the largest output change for the same input variation, to guide decision-making. 

In this setting, the sensitivity of the point estimates, such as the moment functions and the failure probabilities, can be obtained using the partial derivatives of the metrics with respect to (w.r.t) the input distribution parameters. Although the general application of the derivative-based sensitivity analysis can be limited by the difficulty of computing the derivatives, the derivatives w.r.t the input distribution parameters can be more easily evaluated by differentiation inside the expectation operator (c.f. \cref{eq:1,eq:2,eq:3,eq:4} in Section \ref{sec:2} ). This is possible because the individual samples of the random output are not directly dependent on the input distribution parameters.  As a result, the partial derivative operation is only evaluated w.r.t the joint probability density function (PDF) of the input, and this approach is called the likelihood ratio/score function method (LR/SF) \cite{spall_introduction_2003, Rubinstein_2016}. 

As described, the LR/SF method is merely a mathematical trick. Nevertheless, if used together with a sampling method, it is efficient as the uncertainty metric and its sensitivity can be evaluated in a single simulation run (c.f. Section \ref{sec:3.4}). The LR/SF method has been applied to general objective functions in stochastic optimization \cite{spall_introduction_2003}, the failure probability in reliability engineering \cite{li_likelihood_2011} and some distribution-free properties of the LR/SF method are discussed in \cite{millwater_universal_2009}. 

The sensitivity of entropy, on the other hand, cannot be directly evaluated using the LR/SF method. Instead, sensitivity related to entropy is often analysed using the Kullback–Leibler (K-L) divergence (aka relative entropy), by measuring the divergence between two PDFs (probability density functions) corresponding to two different cases. This approach is studied in \cite{PARK1994253} for safety assessment to explore the impact on risk profile due to input uncertainties and in \cite{liu_relative_2005} for engineering design before and after uncertainty reduction of the random variables of interest. A similar approach using the mutual information between the input and the output has also been studied for sensitivity analysis \cite{ludtke_information-theoretic_2008}. The mutual information can be regarded as a special form of the K-L divergence except that it requires the use of the joint PDF. As the K-L divergence is not a metric, alternative distance measures such as the Hellinger distance has been proposed to quantify the difference between two PDFs and the corresponding sensitivities \cite{JIA2014103}. 

It should be noted although the relative entropy is not a metric, its infinitesimal form is directly linked to the Fisher information \cite{Fisher1922On} which is a metric tensor and this link has been explored in \cite{yang_digital_2022} for probabilistic sensitivity analysis using the Fisher information matrix (FIM). The LR/SF method can then be used to compute the FIM efficiently for sensitivity analysis of the output entropy \cite{yang_digital_2022}. 

In this paper, we propose a new sensitivity matrix $\bf r$ that unifies the sensitivity of a wide range of commonly used uncertainty metrics, from moments of the uncertain output to the entropy of the entire distributions, in a single framework. This is made possible by the likelihood ratio/score function method (LR/SF) where the sensitivity to the input distribution parameters of different metrics can be expressed in the same form (c.f. Eq \ref{eq:4}). The 2nd moment of the sensitivity matrix, $\Err$, arises naturally when the impact of input perturbation on the output is examined. Moreover, the maximization of the perturbation of the output uncertainty metrics leads to an eigenvalue problem of the matrix $\Err$. The eigenvalues represent the magnitudes of the sensitivities with respect to (w.r.t) simultaneous variations of the input distribution parameters $\mathbf{b}$, and the relative magnitudes and directions of the variations are given by the corresponding eigenvectors. Therefore, the eigenvectors corresponding to the largest eigenvalues are the most sensitive directions to guide decision-making.

The sensitivity matrix $\bf r$ can be seen as a counterpart of the deterministic sensitivity matrix (Jacobian matrix) as the elements of $\bf r$ are the normalised partial derivatives of the output uncertainty metrics w.r.t to the distribution parameters of the uncertain input (c.f. Eq \ref{eq:6}). The resulting eigenvectors, therefore, have direct sensitivity interpretation. It should be noted that although the sensitivity matrix $\bf r$ is formulated and estimated using the LR/SF method, the use of the 2nd moment matrix and its eigenvectors for sensitivity analysis additionally captures the interactions of the sensitivities of different metrics. 

In addition, the current work is motivated by a recent study \cite{yang_combined_2022} where a special case of the proposed sensitivity matrix has been applied successfully to the combined sensitivity analysis of multiple failure modes. We are going to show that, not only does $\Err$ capture the combined perturbation effect of multiple metrics, e.g., multiple failure modes or multiple moment functions, but also include the Fisher information matrix (FIM) as a special case. Application of the FIM for sensitivity analysis can be found in many areas of science and engineering. For example, the Fisher Information Matrix (FIM) has been applied to the parametric sensitivity study of stochastic biological systems \cite{gunawan_sensitivity_2005}, to assess the most sensitive directions for climate change given a model for the present climate \cite{majda_quantifying_2010} and as one of the process-tailored sensitivity metrics for engineering design \cite{yang_digital_2022}.

It should be noted that there are two main differences between the proposed framework and the commonly used variance-based sensitivity analysis \cite{saltelli_global_2008}. First, variance-based approaches study how the variance of the output can be decomposed into contributions from uncertain inputs. It ranks the factors based on the assumption that the factor can be fixed to its true value, i.e., complete reduction of the uncertainties, which is rarely possible in practice \cite{oakley_probabilistic_2004}. In contrast, the proposed framework uses partial derivatives to examine the perturbation of the output metrics due to a variation of the input distribution parameters. As the distribution parameters are often based on data, it is equivalent to asking which uncertain dataset the decision-makers should focus on to change the output the most. And this is particularly pertinent to data-driven applications like digital twins \cite{yang_digital_2022}. Second, the output sensitivity measure from the variance-based methods is the percentage contribution, of each factor or the interactions between factors, to the output variance. The proposed framework, on the other hand, outputs the eigenvectors of the sensitivity moment matrix $\Err$ as the principal sensitivity directions for simultaneous variations of the input distribution parameters. This is based on a more pragmatic view that given a finite budget to change the parameters, maximizing the impact on the output follows the principal sensitivity directions, which tend to be a simultaneous variation of the parameters because their effects on the output are likely to be correlated. More discussions on the budget constraint can be found in Section \ref{sec:3.2} with a generalization to the generalised eigenvalue problem. 

It should be noted that despite the differences, for some cases, the aggregated index for individual parameters given in Eq \ref{eq:23} can be used to compare the Fisher sensitivity results against the variance-based main and total sensitivity indices. This has been done in \cite{Yang_symplectic} for a 15-dimensional problem, and in that case, the dominant first eigenvector of the FIM seems to correspond to the main effects from the variance-based sensitivity analysis. 

In what follows, the general sensitivity framework is introduced in Section \ref{sec:2} where the sensitivity analysis is reformulated as a standard eigenvalue problem. In Section \ref{sec:3}, we discuss various properties of the proposed framework, including the link to the Fisher information matrix and the possible extension to a generalised eigenvalue problem for robust sensitivity analysis. A benchmark study, using two commonly used functions for sensitivity analysis, is conducted in Section \ref{sec:4} for the Fisher sensitivity. Three different examples are considered in Section \ref{sec:5}, ranging from a simple cantilever beam to an offshore marine riser, to demonstrate the potential applications of the proposed sensitivity framework. Concluding remarks are given in Section \ref{sec:6}. 

\section{Sensitivity framework}
\label{sec:2}

Consider a general function $\bf y=h(x)$, the probabilistic sensitivity analysis characterises the uncertainties of the outputs $\bf y$ that are induced by the random inputs $\bf x$. It is assumed that the uncertainties of $\bf x$ can be described by parametric probability distributions, i.e., $x \sim p(\mathbf{x|b})$, where $\bf b$ are the distribution parameters. 

One commonly used summary statistic is the (central) moment function of the uncertain output, such as the mean and variance. More generally, the moment function is taken with respect to a function of the uncertain output $g(\mathbf y)$. This might arise when there is a stochastic process present, such as the random forces considered in some of the examples in Section \ref{sec:5}, and the $g(\cdot)$ function could represent max, min or root mean square (r.m.s). In this setting, the $q^\text{th}$ moment function and its partial derivative w.r.t the input distribution parameters can be expressed as:
\begin{subequations} \label{eq:1}
    \begin{align}
     m_q  &  =  \mathbb E_X\left[ g^q(\mathbf{h(x)}) \right] = \int g^q(\mathbf{h(x)}) \pxb \dd{\mathbf x}  \label{eq:1a}   \\
     \frac{\partial m_q}{\partial \mathbf b} & = \int g^q(\mathbf{h(x)}) \frac{\partial \pxb}{\partial \mathbf b} \dd{\mathbf x}  \label{eq:1b} 
    \end{align}
\end{subequations}
where it has been assumed that the differential and integral operators are commutative, i.e. the order of the two operations can be exchanged under regularity conditions of continuous and bounded functions. 

Another metric is the probability of failure and its gradient: 
\begin{subequations} \label{eq:2}
    \begin{align}
     P_f & =  \mathbb E_X\left[ \mathrm{H} \left[ g\left( \mathbf{h(x)} \right) - z \right] \right] = \int \mathrm{H} \left[ g\left( \mathbf{h(x)} \right) - z \right] \pxb \dd{\mathbf x} \label{eq:2a}   \\
     \frac{\partial P_f}{\partial \mathbf b} & = \int \mathrm{H} \left[ g\left( \mathbf{h(x)} \right) - z \right] \frac{\partial \pxb}{\partial \mathbf b} \dd{\mathbf x}  \label{eq:2b}
    \end{align}
\end{subequations}
where $\mathrm {H} (\cdot)$ is the Heaviside step function and $z$ represents the failure threshold. It is noted in passing that the application of failure probability is not limited to reliability engineering. For example, the probability of cost-effectiveness in health economics \cite{baio_probabilistic_2015} and the probability of acceptability in design \cite{wallace_design_1996} can both be formulated in the same way as Eq \ref{eq:2a}. 

When the quantity of interest is the underlying distribution function of the uncertain outputs, the density function and its gradient w.r.t the distribution parameters can be expressed as \cite{yang_digital_2022}:  
\begin{subequations} \label{eq:3}
    \begin{align}
     p(\mathbf y) & =  \mathbb E_X\left[ \prod_n  \delta \left[ y_n - h_n(\mathbf x)\right] \right] = \int \prod_n  \delta \left[ y_n - h_n(\mathbf x)\right] \pxb \dd{\mathbf x}   \label{eq:3a} \\
     \frac{\partial p(\mathbf y)}{\partial \mathbf b} & = \int \prod_n  \delta \left[ y_n - h_n(\mathbf x)\right] \frac{\partial \pxb}{\partial \mathbf b} \dd{\mathbf x}  \label{eq:3b} 
    \end{align}
\end{subequations}
where $\delta(\cdot)$ is the Dirac delta function. 

Although the aforementioned diverse metrics measure different aspects of the uncertain output, it is clear that all of them can be more compactly described using a general utility function:
\begin{subequations} \label{eq:4}
    \begin{align}
     U & =  \mathbb E_X\left[ u\left( \mathbf x \right) \right] = \int  u\left( \mathbf x \right)  \pxb \dd{\mathbf x}   \label{eq:4a} \\
     \frac{\partial U}{\partial \mathbf b} & = \int u\left( \mathbf x \right)  \frac{\partial \pxb}{\partial \mathbf b} \dd{\mathbf x}  \label{eq:4b} 
    \end{align}
\end{subequations}
where the utility function $u(\mathbf x)$ represents the $g^q(\cdot)$ in Eq \ref{eq:1}, $\mathrm {H} (\cdot)$ in Eq \ref{eq:2} and $\delta(\cdot)$ in Eq \ref{eq:3}. It should be noted that the utility function could also depend on other variables, such as the failure threshold $z$ for the case of failure probability. However, $u(\mathbf x)$ is not directly dependent on the parameters $\bf b$, as $\mathbf {b} \rightarrow \mathbf{x} \rightarrow u(\mathbf x)$ forms a Markov chain.  As a result, it is possible to differentiate the joint PDF $p(\mathbf{x|b})$ within the integral in Eq \ref{eq:4b}. And that is the same for \cref{eq:1,eq:2,eq:3}. As mentioned in the introduction, this approach is sometimes called the likelihood ratio/score function method (LR/SF). An advantage of this approach is that, if used together with a sampling method such as the Monte Carlo method, the uncertainty quantification and sensitivity analysis can be conducted in a single simulation run and more details are given in Section \ref{sec:3.4}.  

The purpose of our sensitivity analysis is to identify the most important uncertain parameters, i.e., which set of parameters would perturb the output of interest the most. This perturbation can be quantified as $\Omega = (\Delta U/U)^2$, where the normalisation leads to percentage perturbation and the square operation quantifies the absolute value. If a first-order perturbation is assumed, a general form of the normalised perturbation is: 
\begin{equation} \label{eq:5}
   \begin{split}
       \Omega &  = \mathbb E \left[ \sum_k \left( \frac {\Delta U_k}{U_k} \right)^2 \right]  \\
                    &  =  \mathbb E \left[ \sum_k \left( \frac{1}{U_k} \frac {\partial U_k}{\partial \mathbf {b}} \Delb\right)^2 \right]  \\
                    &  = \sum_i\sum_j\Delta b_i \Delta b_j \mathbb E \left[  \sum_k r_{ik} r_{jk}  \right] \\
                     & = \Delb^{\T} \Err \Delb
  \end{split}
\end{equation}
where the $jk^{\text{th}}$ entry of the matrix $\mathbf r$ is defined accordingly as:
\begin{equation} \label{eq:6}
      r_{jk}  =  \frac{1}{U_k} \frac {\partial U_k}{\partial b_j}
\end{equation}
The matrix $\bf r$ can be seen as a counterpart of the deterministic sensitivity matrix (Jacobian matrix) and therefore called sensitivity matrix in this paper. It is interesting to note that the 2nd moment of the sensitivity matrix, $\Err$, arises naturally from the perturbation analysis. As it is in the form of a Gram matrix, $\Err$ is symmetric positive semi-definite (also evident from the quadratic form of Eq \ref{eq:5}). 

The general form of the perturbation in Eq \ref{eq:5} considers the combined effect of multiple utilities. For example, there could be multiple failure modes where $U_k = P_f^{(k)}$ denotes the $k^{\text{th}}$ failure mode; it is also often of interest to consider the combined sensitivity of multiple responses or moments of the same uncertain output where $U_k = m_k$ denotes the $k^{\text{th}}$ moment. It is noted in passing that a weighting could be added to each $U_k$ and that would result in a weighting of $r_{jk}$ in Eq \ref{eq:6}. The weighted scenario is not considered further in this paper as the weighting is strongly case-dependent but will not alter the general form of Eq \ref{eq:5}. The expectation operation $\mathbb E [\cdot]$ in Eq \ref{eq:5} takes account of any additional uncertainties that might arise in different cases. For example, the failure threshold $z$ could be uncertain in Eq \ref{eq:2}; for the case of the joint density function, where  $U = p(\mathbf y) $, the gradient of the log utility described in Eq \ref{eq:6} is uncertain due to randomness of the output $\bf y$. 

Using the general perturbation function described in Eq \ref{eq:5}, the sensitivity analysis can be formulated as a constrained optimization problem:
\begin{equation} \label{eq:7}
\begin{split}
          \text{max} &  \quad  \frac{1}{2}\Omega = \frac{1}{2}\Delb^{\T} \Err \Delb     \\
          \text{s.t.}  & \quad\Delb^{\T}\Delb = \epsilon
\end{split}
\end{equation}
where the method of Lagrange Multiplier can be used:
\begin{subequations} \label{eq:8}
    \begin{align}
     L & = \Delb^{\T} \Err \Delb - \lambda(\Delb^{\T}\Delb - \epsilon)   \label{eq:8a} \\
     \frac{\partial L}{\partial \Delb}  
     & = \Err \Delb -  \lambda \Delb \label{eq:8b} 
    \end{align}
\end{subequations}
where $\lambda$ is the Lagrange multiplier. Setting the first order optimality condition for the Lagrangian, Eq \ref{eq:8b} then leads to the following standard eigenvalue problem:
\begin{equation} \label{eq:9}
    \Err \mathbf {q} = \lambda \mathbf {q}
\end{equation}
The eigenvalues represent the magnitudes of the sensitivities with respect to (w.r.t) simultaneous variations of the parameters $\mathbf{b}$, and the relative magnitudes and directions of the variations are given by the corresponding eigenvectors. As the solution to a maximization problem, the eigenvectors corresponding to the largest eigenvalues then provide the most perturbation of $\Omega$ in Eq \ref{eq:5}. 
\section{Discussion}
\label{sec:3}

\subsection{Information theoretical metrics as a special case}
\label{sec:3.1}
When the utility $U$ corresponds to a probably or probability density, the expression for $\bf r$ in Eq \ref{eq:6} can be more compactly written as:
\begin{equation} \label{eq:10}
      r_{jk}  =  \frac{\partial \log U_k}{\partial b_j}
\end{equation}
As the log probability can be seen as the information content of a random event in information theory, the expression in Eq \ref{eq:10} reveals the information link of the sensitivity framework. 

More concretely, for the case described in Eq \ref{eq:3},  the utility corresponds to $p(\mathbf{y})$ and the resulted perturbation is (using \cref{eq:5,eq:6}):
\begin{equation} \label{eq:11}
   \begin{split}
       \Omega  &  =  \mathbb E_Y\left[ \left( \frac {\Delta \py}{\py} \right)^2\right]    \\
                       &   =  \Delb^{\T}\mathbb E_Y\left[ \frac { \partial \log \py}{\partial \mathbf{b}} {\frac {\partial \log \py}{\partial \mathbf{b}}}^{\T} \right]  \Delb
    \end{split}    
\end{equation}
where the expectation part in the right hand side of the equation can be more explicitly written as: 
\begin{equation} \label{eq:12}
   \begin{split}
       \mathbb{E}_Y\left[ \frac { \partial \log \py}{\partial \mathbf{b}} {\frac {\partial \log \py}{\partial \mathbf{b}}}^{\T} \right] 
        &  =  \int  \frac { \partial \log \py}{\partial \mathbf{b}} {\frac {\partial \log \py}{\partial \mathbf{b}}}^{\T}  \py \dd{\mathbf y}    \\
        &  = \int  \frac { \partial \py}{\partial \mathbf{b}} {\frac {\partial  \py}{\partial \mathbf{b}}}^{\T} \frac{1}{ \py} \dd{\mathbf y} 
    \end{split}
\end{equation}
and this is the Fisher information matrix (FIM) \cite{cover_elements_2006} and is denoted as $\mathbf F$. Therefore, the perturbation in Eq \ref{eq:11} can be rewritten as:
\begin{equation} \label{eq:13}
       \Omega  =  \Delb^{\T} \mathbf{F}  \Delb \approx  2\text{KL} \left[ \pyb || \pybb \right]
\end{equation}
where $\text{KL} [\cdot]$ indicates Kullback-Leibler (K-L) divergence and it is also called relative entropy \cite{cover_elements_2006}.  The approximation in Eq \ref{eq:13} can be found via a Taylor expansion of the perturbed PDF $\pybb$ from the relative entropy expression, with the third and higher order terms ignored. It should be noted that Eq \ref{eq:13} has been derived in \cite{yang_digital_2022} to link the relative entropy and the FIM. However, in this paper, we have extended that link to a general perturbation metric $\Omega$ which unifies several other metrics as well. 

In Section \ref{sec:4}, a benchmark study is conducted for the Fisher sensitivity. As mentioned in the introduction, the application of the Fisher information for sensitivity analysis can be found in many areas of science and engineering. In the numerical examples given in Section \ref{sec:5}, we will demonstrate that the FIM can be utilised in different ways using the proposed framework. 
\subsection{Extension to generalised eigenvalue problems}
\label{sec:3.2}
The constraint in Eq \ref{eq:7} controls the potential change of the parameters. This limit can be seen as a result of the finiteness of resources which is true for all physical systems. A more general decision-oriented constraint can be written as:
\begin{equation} \label{eq:14}
       \Delb^{\T} \mathbf{W} \Delb = \epsilon
\end{equation}
where $\mathbf{W}$ is a weighting matrix which is symmetric. Instead of the standard eigenvalue equation in Eq \ref{eq:9}, the weighted constraint from Eq \ref{eq:14} leads to a generalised eigenvalue problem:
\begin{equation} \label{eq:15}
    \Err \mathbf {q} =  \lambda  \mathbf{W} \mathbf {q}
\end{equation}
Consider a scenario where the interest is to understand the sensitivity of the failure probability. Eq \ref{eq:9} then leads us to the standard eigenvalue analysis of the following matrix:
\begin{equation} \label{eq:16}
     \mathbb E \left[ \mathbf{rr}^{\T} \right]  =  \mathbb E \left[ \frac { \partial \log P_f}{\partial \mathbf{b}} {\frac {\partial \log P_f}{\partial \mathbf{b}}}^{\T}\right]  
\end{equation}
where the expectation is w.r.t the potential uncertain failure threshold. Eq \ref{eq:16} is rank-1 for a deterministic threshold. However, the resulting sensitivity directions for $\Delb$ might impact both the safe and failure regions. One way to mitigate the unwanted perturbation is to set an uncertainty constraint using the relative entropy:
\begin{equation} \label{eq:17}
\begin{split}
          \text{max} &  \quad  \frac{1}{2}\Omega = \frac{1}{2}\Delb^{\T}  \mathbb E \left[ \frac { \partial \log P_f}{\partial \mathbf{b}} {\frac {\partial \log P_f}{\partial \mathbf{b}}}^{\T}\right]  \Delb     \\
          \text{s.t.}  & \quad \text{KL} \left[ \pyb || \pybb \right] = \epsilon
\end{split}
\end{equation}
and this leads to the following generalised eigenvalue problem (with the substitution of Eq \ref{eq:13} for the approximation of the relative entropy):
\begin{equation} \label{eq:18}
     \mathbb E \left[ \frac { \partial \log P_f}{\partial \mathbf{b}} {\frac {\partial \log P_f}{\partial \mathbf{b}}}^{\T}\right]   \mathbf {q} 
     =  \lambda  \mathbf{F} \mathbf {q}
\end{equation}
where the Fisher information matrix $\bf F$ now takes the position of $\bf W$ in Eq \ref{eq:15}. Eq \ref{eq:18} will be utilised in one of the numerical examples to demonstrate the application for constrained perturbation of a failure probability. 
\subsection{Reparameterization and normalisation}
\label{sec:3.3}
The matrix $\Err$ depends on the parametrization used. Suppose $b_j=\phi_j(\theta_i), \: i=1,2,\dots,s$,  then the $k^{\text{th}}$ column of the sensitivity matrix $\mathbf r$ with respect to $\bm{\uptheta}$ is:
\begin{equation} \label{eq:19}
      \mathbf{r}_{k}  =  \frac{1}{U_k} \frac {\partial U_k}{\partial \bm{\uptheta}} 
                 = \mathbb{J}^{\T}\frac{1}{U_k} \frac {\partial U_k}{\partial \mathbf{b}} 
\end{equation}
where $\mathbb{J}$ is the Jacobian matrix with $\mathbb{J}_{ji}={\partial b_j}/{\partial \theta_i}$.
As a result, the matrix $\Err$ w.r.t the parameters $\bm{\uptheta}$ can be found via a reparameterization: 
\begin{equation} \label{eq:20}
    \Err_{\bm{\uptheta}}
    =\mathbb{J}^{\T}\Err_{\mathbf{b}}\mathbb{J} 
\end{equation}
Normalization w.r.t the parameters $\bf b$ is equivalent to a reparametrization. In the case of proportional normalization, where the sensitivity matrix becomes dimensionless, the change of parameter is $b_j=\bar{b}_j\theta_j$ with $\bar{b}_j$ the nominal value for normalization, and the Jacobian matrix in Eq \ref{eq:20} is just a diagonal matrix with $\bar{b}_j$ on the diagonal. Due to its simplicity for sensitivity analysis, proportional normalization is applied to the numerical cases considered in Section \ref{sec:5}. 
\subsection{Numerical considerations}
\label{sec:3.4}
A general function $\bf y=h(x)$ can rarely be solved analytically. The unique mathematical form of \cref{eq:1,eq:2,eq:3}, and more generally Eq \ref{eq:4}, allows efficient computation of the gradient if a sampling method such as Monte Carlo (MC) method is used.  The Monte Carlo approximation of the integrals in Eq \ref{eq:4} results:
\begin{subequations} \label{eq:21}
    \begin{align}
    \begin{split}
     U & = \int  u\left( \mathbf x \right)  \pxb \dd{\mathbf x}   \\
     &  \approx \frac{1}{N} \sum_i  u\left( \mathbf x_i \right)  
    \end{split}     
      \label{eq:21a} \\
    \begin{split} 
     \frac{\partial U}{\partial \mathbf b} & 
      = \int u\left( \mathbf x \right)  \frac{\partial \log \pxb}{\partial \mathbf b} \pxb \dd{\mathbf x} \\
    &  \approx  \frac{1}{N} \sum_i u\left( \mathbf{x}_i \right)  \frac{\partial \log p(\mathbf{x}_i|\mathbf{b})}{\partial \mathbf b}   
    \end{split}      
      \label{eq:21b} 
   \end{align}
\end{subequations}
where $\mathbf{x}_i$ is a MC realisation of the random variable $\bf x$ and $N$ MC simulations are considered. \\
For many commonly used distributions, analytical closed-form expressions can be obtained for the partial derivatives with respect to (w.r.t) a distribution parameter. For example, for a Gaussian distribution:
\begin{subequations} \label{eq:22}
    \begin{align}
     \frac{\partial \log p(x|\mu,\sigma)}{\partial \mu} 
     & = \frac{x-\mu}{\sigma}
     \label{eq:22a} \\
     \frac{\partial \log p(x|\mu,\sigma)}{\partial \sigma} 
     & = \frac{(x-\mu)^2-\sigma^2}{\sigma^3}
      \label{eq:22b} 
   \end{align}
\end{subequations}
where $\mu$ and $\sigma$ are the mean and standard deviation of the Gaussian PDF. Therefore, with the analytical expressions available, the utility of interest and its gradient in Eq \ref{eq:21} can be obtained in a single computational run. This is one of the main numerical advantages of the adopted likelihood ratio/score function method (LR/SF) as described in the introduction. The application of Eq \ref{eq:21} has been validated in \cite{yang_combined_2022} for a failure probability sensitivity, where the perturbation approximated using Eq \ref{eq:21b} agrees well with the exact results from a direct perturbation. In \cite{yang_digital_2022}, the Fisher information estimated from Eq \ref{eq:21} also identifies the influential parameters for a marine riser as expected. Not only is it efficient, but it is also clear that Eq \ref{eq:21} is independent of the parameter dimensions and thus the sensitivity approach is not limited by the dimension of the input variables. It should be noted that although the sensitivity framework introduced applies to dependent inputs, it is assumed for simplicity in the numerical implementation below that the components of $\bf x$ are independent. 

As the numerical computation is based on the LR/SF method, the sensitivity estimation is independent of the dimension of the input parameters \cite{Rubinstein_2016}. This is in contrast to the variance-based methods where the computational cost is proportional to the input dimension \cite{saltelli_global_2008}. However, a large number of sampling points might be required to maintain a low variance from the LR/SF estimation. As our main purpose is to present the new sensitivity framework, only the standard Monte Carlo (MC) method is used. Advanced simulation methods, such as importance sampling and subset simulation, can be used to improve the sampling efficiency. 

Based on the LR/SF method estimation, the proposed framework requires an eigen-decomposition of  the $\Err$ matrix. As the size of the 2nd moment matrix depends on the dimension of the input parameters, for high dimension problems, an iterative approach might be needed to solve the eigenvalue decomposition. Nevertheless, as the 2nd moment matrix is symmetric and it is often the first few largest eigenvalues that are of interest for sensitivity analysis, additional cost for the eigenvalue problem should be minimal compared to the MC sampling. As a result, the computational cost of our method is similar to the LR/SF method in general. 
\section{Benchmark study}
\label{sec:4}
The Fisher information has been introduced in \cite{yang_digital_2022} for sensitivity analysis with respect to the input distribution parameters. The examples given in \cite{yang_digital_2022} have focused on the demonstration for engineering design applications, but lack benchmarking against commonly used functions for sensitivity analysis. As the Fisher sensitivity will be used throughout the examples in Section \ref{sec:5}, in this section, we present benchmark studies against two of the commonly used variable screening benchmark functions. For screening purposes, the input variables for both functions are assumed to be from a standard Gaussian distribution, i.e., $x \sim \mathbb{N}(0, 1)$. 

\subsection{Decreasing coefficient function}
\label{sec:4.1}
The first benchmark function is a linear function with decreasing coefficients:
\begin{equation} \label{eq:dcf}
      f(\mathbf x) = 0.2x_1 + \frac{0.2}{2}x_2 +  \frac{0.2}{4}x_3 + \frac{0.2}{8}x_4  
                                + \frac{0.2}{16}x_5 + \frac{0.2}{32}x_6 +\frac{0.2}{64}x_7 +\frac{0.2}{128}x_8 +  e                                
\end{equation}
where $e \sim \mathbb{N}(0, 0.05^2)$ is a small error term. This function has been used for input variable screening in \cite{Linkletter_2006}, where it concluded that the first three variables, $x_1, \ x_2, \ \text{and} \ x_3$, are more important than the rest.  

\begin{figure}[!h]
	\centering
	 \includegraphics[width=12cm]{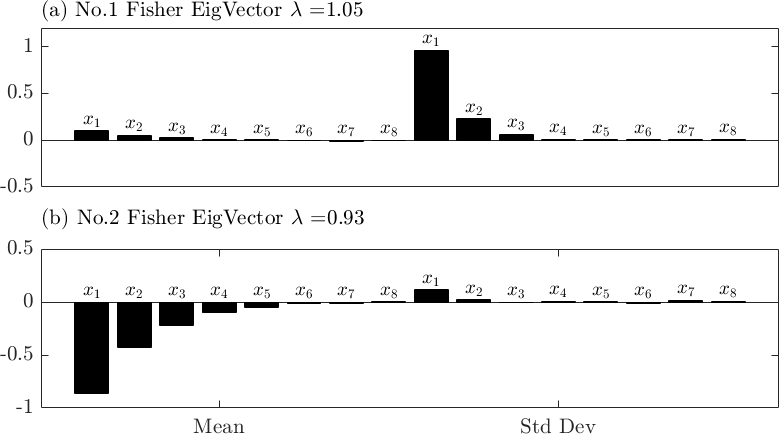}  
	\caption{Fisher sensitivity results for the decreasing coefficient function described in Eq \ref{eq:dcf}. }
	\label{fig:dcf1}
\end{figure}

The sensitivity analysis using the Fisher information matrix (FIM) from Eq \ref{eq:12} is applied to the function above and the results are presented in Figure \ref{fig:dcf1}. In contrast to variance-based sensitivity approaches, as explained in the introduction, the proposed framework provides the principal sensitivity directions w.r.t the distribution parameters, e.g., the mean and standard deviation (Std Dev) in Figure \ref{fig:dcf1}.

\begin{figure}[!h]
	\centering
	 \includegraphics[width=10cm]{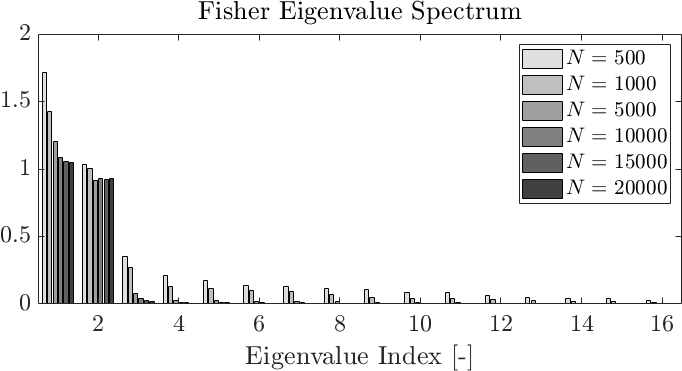}  
	\caption{The eigenvalue spectrum of the Fisher information matrix, for the decreasing coefficient function described in Eq \ref{eq:dcf}. }
	\label{fig:dcf2}
\end{figure}

The sensitivity results in Figure \ref{fig:dcf1} display the first two eigenvectors of the FIM, as the rest of the eigenvalues have negligible amplitudes as shown in Figure \ref{fig:dcf2}. The results in Figure \ref{fig:dcf1} clearly show that the importance of the variables decreases as the subscript index increases. It is clear that the first three variables are more important than the rest as seen from the 1st eigenvector, and that agrees with the conclusion from \cite{Linkletter_2006}. Moreover, for the 1st eigenvector shown in Figure \ref{fig:dcf1}(a), the ratio between the sensitivities to the standard deviations of the first three variables are $[x_1:x_2:x_3] = [1:0.236:0.065]$. This ratio is very close to the ratio of the square of the coefficients, i.e., $[1:1/4:1/16]$. Similarly, the dominant sensitivities to the mean of the first three variables, as given in Figure \ref{fig:dcf1}(b), have the ratio of $[1:0.496:0.260]$. This suggests that not only does the FIM sensitivity index give the correct ranking of the variables, but it has also accurately captured the relative strength of the linear coefficients of Eq \ref{eq:dcf}.   

In Figure \ref{fig:dcf2}, the spectrum of the FIM eigenvalues is given, for different numbers of Monte Carlo samples (c.f. Section \ref{sec:3.4}). This serves as a convergence test for the FIM estimation. The same type of convergence tests has also been performed for the numerical examples in Section \ref{sec:5} but not presented for conciseness.   

\subsection{Roos \& Arnold Function}
\label{sec:4.2}
\begin{figure}[!h]
	\centering
	 \includegraphics[width=10cm]{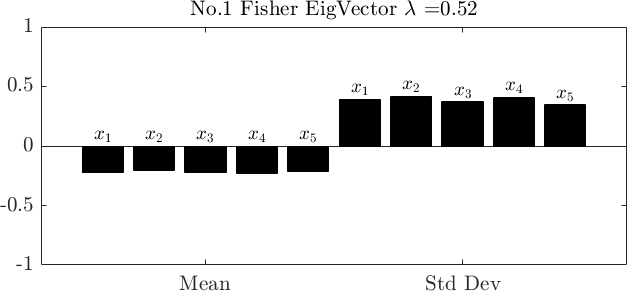}  
	\caption{Fisher sensitivity results for the Roos \& Arnold function described in Eq \ref{eq:raf}. }
	\label{fig:raf1}
\end{figure}

\begin{figure}[!h]
	\centering
	 \includegraphics[width=10cm]{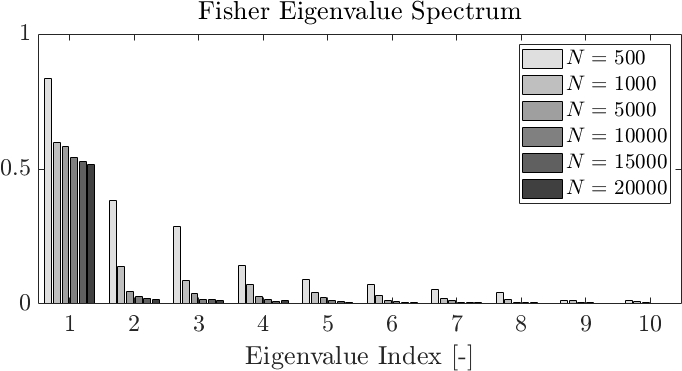}  
	\caption{The eigenvalue spectrum of the Fisher information matrix, for the Roos \& Arnold function described in Eq \ref{eq:raf}. }
	\label{fig:raf2}
\end{figure}

The second benchmark function is a product function:
\begin{equation} \label{eq:raf}
      f(\mathbf x) = \prod_{i=1}^d \vert 4x_i - 2 \vert                                
\end{equation}
This function occurs as an integrand multiple times in the literature. It has been classified as a Type C function in \cite{Kucherenko_2011}, meaning that it has dominant high-order interaction terms. It is clear from Eq \ref{eq:raf} that there is no difference between the input variables. The results from the FIM, similar to the previous example, are shown in Figure \ref{fig:raf1} with five input variables. In this case, the first eigenvalue is much bigger than the rest as shown in Figure \ref{fig:raf2} (which also shows the convergence results as previous example). The sensitivity results in Figure \ref{fig:raf1} agree well with expectations that the input variables are of similar importance. It should be noted that the amplitudes of the five variables shown in Figure \ref{fig:raf1} are not exactly the same from the Fisher sensitivity. This is because the FIM is based on a sampling approach and the samples for the five input variables will not be exactly the same as they are independently sampled. 

\section{Application examples}
\label{sec:5}
\begin{figure}[!h]
	\centering
	 \subfloat[\centering Case 1] {{\includegraphics[width=6cm]{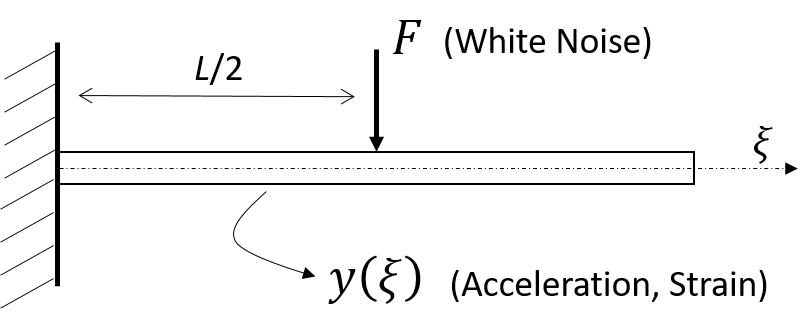} } \label{fig:1a}} 
	 \hfill
    \subfloat[\centering Case 2]{{\includegraphics[width=4cm]{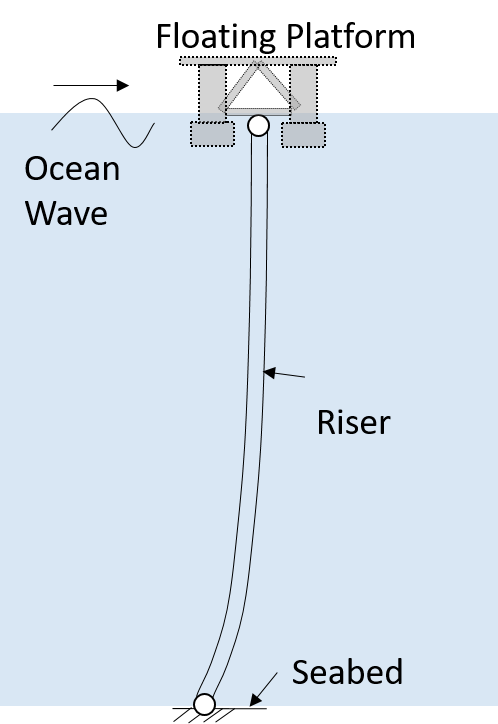} }\label{fig:1b}}
    	 \hfill
	  \subfloat[\centering Case 3]{{\includegraphics[width=3cm]{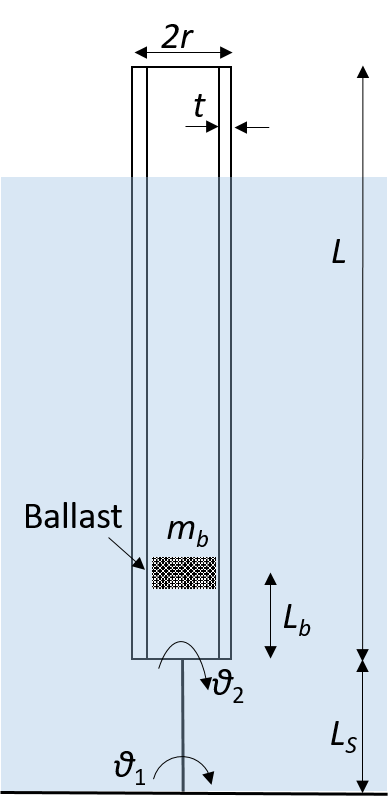} }\label{fig:1c}}
	\caption{Three demonstrating examples. (a) Case 1, a cantilever beam with white noise excitation; (b) Case 2, an offshore marine riser subject to random wave loading; (c) Case 3, a model floating column in a wave tank.}
	\label{fig:1}
\end{figure}
\subsection{Combined sensitivity for multiple responses}
\label{sec:5.1}
In this section, we use a simple cantilever beam as an example to demonstrate the combined sensitivity analysis for multiple responses using the proposed sensitivity framework. The cantilever beam, case 1 in Figure \ref{fig:1}, is subject to a white noise excitation of unit amplitude. The frequency response functions for both acceleration and strain responses, at different positions along the beam, are obtained via modal summation. To keep it simple, only the first mode is considered with a constant modal damping of 0.1. The linear vibration equation of a cantilever beam can be found in many vibration/mechanics textbooks, see e.g., \cite{meirovitch1986elements}, and it is therefore not repeated here for conciseness. The code for this example study can be found in the address given in \nameref{sec:data}. 

The quantities of interest for our sensitivity analysis are the peak r.m.s acceleration and strain responses. ‘peak’ indicates the maximum response along the beam for each sample of the random inputs. The $g(\cdot)$ function in Eq \ref{eq:1} is thus the composition of the r.m.s and maximum functions in this case. The two types of response are normalised by the maximum values across the ensemble of the random samples, i.e., the peak r.m.s results are between 0 and 1. 
\begin{table}
    \footnotesize
	\caption{Mean ($\mu$) and Coefficient of Variation (CoV) for the random variables. }
	\centering
\makebox[\textwidth][c] {
	\begin{tabular}{cccccc}
		\toprule
		             & Young’s Modulus	& Density &	Length	 & Width &	Thickness \\
		\cmidrule(lr){2-6}
	               & $E [Pa]$	&  $\rho [kg/m^3]$	  & $L [m]$	  & $w [m]$	 & $t [m]$ \\
		\cmidrule(lr){2-6}
		    Mean  &  	69e9	 & 2700	& 0.45 &	2e-2	&  2e-3       \\
	  \cmidrule(lr){2-6}   
	          Scenario-A    &	\multicolumn{5}{c}{\textbf{Normal} distribution with CoV = 0.1 }                   \\
		\cmidrule(lr){2-6}  	
		      Scenario-B     & \multicolumn{5}{c}{ \textbf{Gamma} distribution with CoV = 0.5 }                   \\
		\bottomrule
	\end{tabular}
	}
	\label{tab:1}
\end{table}
\begin{figure}[!h]
	\centering
	 \includegraphics[width=12cm]{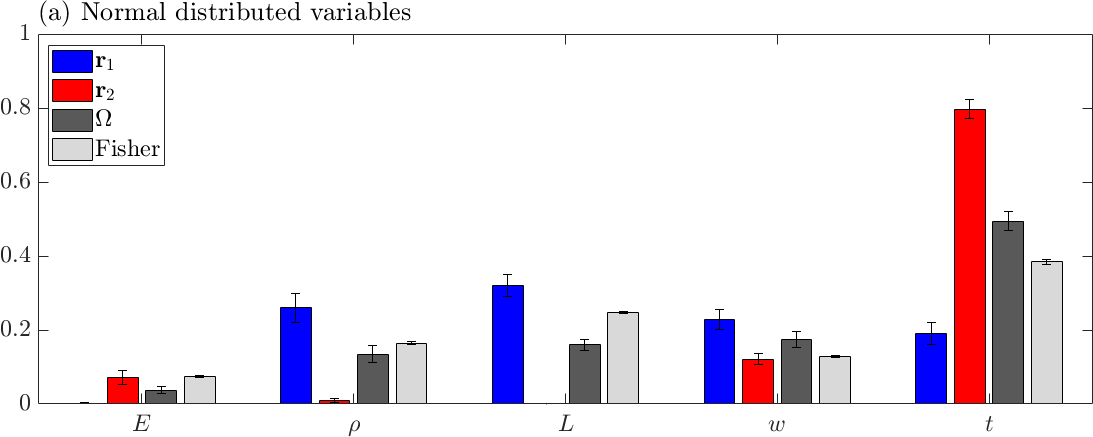}  \quad
	 \includegraphics[width=12cm]{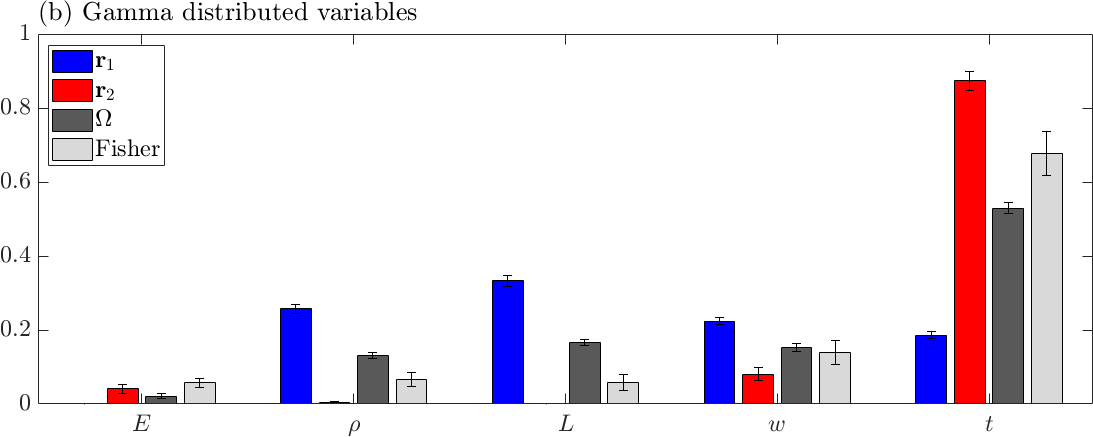} 
	\caption{Variable ranking for multiple responses of the cantilever beam in Figure \ref{fig:1a}. $\mathbf{r_1}$, $\mathbf{r_2}$ indicate the sensitivity of the $U_1$ (acceleration) and $U_2$ (strain) respectively. $\Omega$ is for the combined sensitivity results for $U_1$ and $U_2$ using the metric proposed in this paper, and `Fisher' denotes the sensitivity results from the FIM using the jPDF of the responses. Error bars indicate the standard deviation from 10 repetitions of estimation with 20000 samples for each run. (a) Scenario-A with Normal random input; (b) Scenario-B with Gamma random input.}
	\label{fig:2}
\end{figure}

The parameters for the random variable are listed in Table \ref{tab:1}. Two scenarios are considered with the same mean value but different types of distributions, one with \textit{Normal} (Gaussian) distribution and the other one with \textit{Gamma} distribution, of which the coefficients of variation (CoV) are different. The sensitivity results are shown in Figure \ref{fig:2} for the relative ranking of the five input random variables. One measure for the relative importance of the $j^\text{th}$ variable can be obtained as:
\begin{equation} \label{eq:23}
      s_j^2 = \sum_i \lambda_i q_{ji}^2 
\end{equation}
where $\lambda$ and $\bf q$ are the eigenvalue and eigenvector of the sensitivity matrix from Eq \ref{eq:9}. 

Eq \ref{eq:23} takes a Pythagorean view to estimate the contributions, in analogy to principal component analysis, within each eigenvector and across different eigenvectors using the eigenvalue amplitudes. However, this summary index from Eq \ref{eq:23} assumes that different principal sensitivity directions, represented by the eigenvectors, can be varied at the same time for the parameters. This aggregated view of importance measure essentially only considers the diagonal entries of the moment matrix $\Err$, as shown in Appendix \ref{appendix:a}. It therefore ignores the interactions between the parameters (off-diagonal terms). Furthermore, the phase information of the sensitivity vectors, i.e., increase or decrease, are also lost using this summary index from Eq \ref{eq:23}. This summary sensitivity index is used only in this case to give a better indication that the proposed sensitivity metric does account for the combined effect, as a direct comparison between eigenvectors is difficult between two different metrics. As the purpose of the sensitivity framework is to find the influential set of distribution parameters, as discussed in the introduction, sensitivity results from the eigenvectors will be used for the next two examples. 

In this case study, the utility function to form $\bf r$ in Eq \ref{eq:6} is the mean value of the uncertain response (the peak r.m.s response for each sample of the random input), i.e., $U_1$ and $U_2 $ are the averaged peak r.m.s acceleration and strain responses respectively. $\mathbf{r}_1$ and $\mathbf{r}_2$ are the normalised LR/SF sensitivity vectors of the two corresponding utility functions as in Eq \ref{eq:6}. 

In Figure \ref{fig:2}, the combined sensitivity result is indicated by $\Omega$. This is computed using Eq \ref{eq:23} with the eigenvalues/eigenvectors of the $\Err$ matrix from Eq \ref{eq:9}, where $\mathbf{r}_1$ and $\mathbf{r}_2$ are the intermediate results for forming the $\Err$ matrix in this case. 

In comparison, the sensitivity results using the Fisher information matrix (FIM) form Eq \ref{eq:12} is also presented in Figure \ref{fig:2}. It can be seen in Figure \ref{fig:2} that $\mathbf{r}_1$ ranks $L$ as more important than $t$, while $\rho$ and $L$ are of little importance from $\mathbf{r}_2$. In contrast, although the absolute agreement varies, the resulting relative ranking of the random variables are very similar between the combined analysis ($\Omega$) and the FIM. The conclusion is similar between scenario-A and scenario-B.     

The FIM results are based on the joint probability density function of the acceleration and strain responses. This is in contrast to the combined sensitivity analysis where the two types of responses are assumed to be independent. Nevertheless, the comparison in Figure \ref{fig:2} demonstrates that the combined analysis takes account of the combined perturbation of multiple responses, as indicated by the derivation in Eq \ref{eq:5}. This is in line with the findings in \cite{yang_combined_2022} for multiple correlated failure modes, which can be seen as a special case of the proposed sensitivity framework in this paper. 
\subsection{Robust failure sensitivity}
\label{sec:5.2}
\begin{table}
    \footnotesize
	\caption{Mean ($\mu$) and Coefficient of Variation (CoV) for the random variables of Case 2}
	\centering
\makebox[\textwidth][c] {
	\begin{tabular}{ccccccccc}
		\toprule
		           &  \multicolumn{1}{p{2cm}}{\centering Morison's added \\ mass coefficient}
		           &  \multicolumn{1}{p{2cm}}{\centering Morison's  \\ drag coefficient}
		            &\multicolumn{1}{p{1.5cm}}{ \centering Riser \\ material density}
		            &\multicolumn{1}{p{1cm}} { \centering  Young’s \\ modulus}
		            & \multicolumn{1}{p{1cm}} { \centering  Oil \\ density}
		             & \multicolumn{1}{p{1cm}} { \centering  Top \\ tension} 
		             &	\multicolumn{2}{p{2cm}} { \centering Material S-N \\ curve coefficients} \\
		\cmidrule(lr){2-9}
	               & $C_a$ [-]  & $C_d$ [-] &  $\rho$  [kg/m3] &  $E$ [GPa]	  & $\rho_0$ 	 [kg/m3]   & $T_0$ [kN]	 & $\alpha$ [GPa] & $\delta$ [-] \\
		\cmidrule(lr){2-9}
		     Mean &  1.5 & 1.1 & 7840 & 200 & 920 & 4905 & 199 & 3       \\
	  \cmidrule(lr){2-9}            
			CoV 	& 0.20 & 0.20 & 0.05 & 0.05 & 0.10 & 0.10 & 0.10 & 0.10   \\
		\bottomrule
	\end{tabular}
	}
	\label{tab:2}
\end{table}
In this section, the generalised eigenvalue problem introduced in Section \ref{sec:3.2} is considered. In particular, we use Eq \ref{eq:18} to demonstrate the possibility to perturb a failure probability with an entropic constraint. This might arise in practice to avoid unwanted perturbation of the system responses, e.g., no perturbation of the response in the safe region, and it can be regarded as a robust failure sensitivity analysis. The example system considered is an offshore marine rise shown in Figure \ref{fig:1b} (Case 2) that is subject to a random wave excitation and the nonlinear wave structure interaction is included in the model. The simulation model for this case study, and Case 3 below, has been developed using the CHAOS hydrodynamic code \cite{yang_code_2022} which uses the semi-empirical Morison's equation \cite{sarpkaya_2010} to estimate wave forces. This case study is taken from \cite{yang_digital_2022} where the details of this model and its sensitivities can be found. Different from \cite{yang_digital_2022} where the failure sensitivity is compared to the Fisher sensitivity results, an entropic constraint is applied to the failure sensitivity in this paper. The code for this example study can be found in the address given in \nameref{sec:data}. 
\begin{figure}[!h]
	\centering
	\includegraphics[width=12cm]{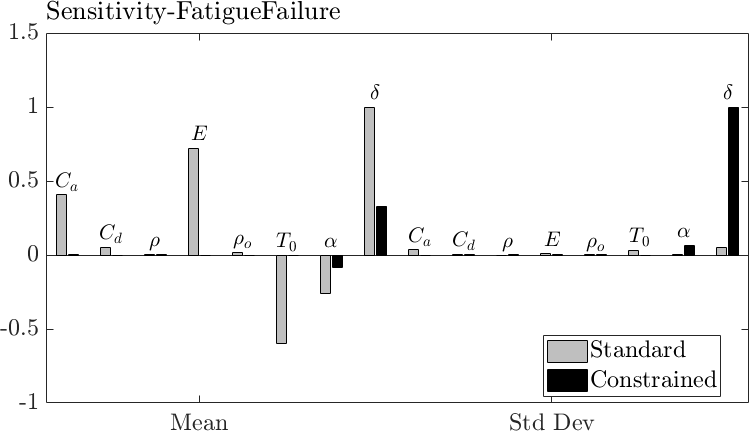}
	\caption{Constrained sensitivity of the failure probability due to fatigue for Case 2. In comparison, the standard sensitivity results of the fatigue failure is also shown.}
	\label{fig:3}
\end{figure}

The parameters for the uncertain variables of Case 2 are listed in Table \ref{tab:2} and the corresponding failure sensitivity results are given in Figure \ref{fig:3}. The results marked as `Constrained' are obtained from solving the generalised eigenvalue equation in Eq \ref{eq:18}. The corresponding results of the standard failure sensitivity vector, which is equivalent to substituting an identity matrix for the matrix $\bf F$  in Eq \ref{eq:18}, is shown in comparison and is denoted as `Standard'. Note that the standard results are the same as in \cite{yang_digital_2022}. 

As can be seen in Figure \ref{fig:3}, the entropy-constrained sensitivity is completely dominated by the S-N coefficients $\alpha$ and $\delta$. This is because the FIM is not dependent on the S-N coefficients that are specific to the fatigue failure analysis. Although a relatively extreme example, this case study demonstrates the potential application of the entropy-constrained sensitivity analysis. It also shows the general applicability of the proposed sensitivity framework once formulated as a generalised eigenvalue problem.
\subsection{Limiting approximation for deterministic inputs}
\label{sec:5.3}
\begin{figure}[!h]
	\centering
	\includegraphics[width=12cm]{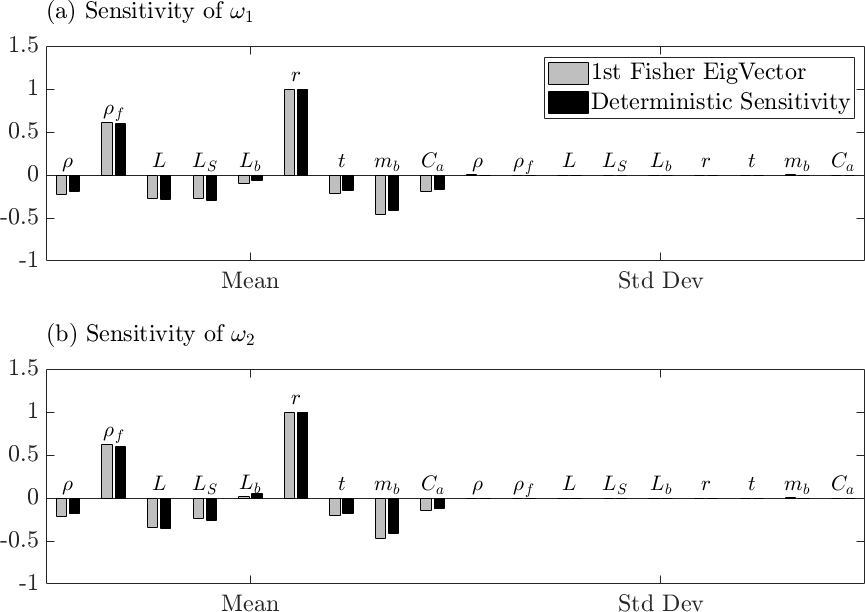}
	\caption{Sensitivity of the two natural frequencies, with a comparison between the dominant Fisher eigenvector and the sensitivity obtained from the analytical approach. The nominal values of the input parameters are used for the deterministic case, and as the mean values for the random case. (a) 1st natural frequency $\omega_1$; (b) 2nd natural frequency $\omega_2$}
	\label{fig:4}
\end{figure}
\begin{table}
    \footnotesize
	\caption{Mean ($\mu$) and Coefficient of Variation (CoV) for the random variables of Case 3}
	\centering
\makebox[\textwidth][c] {
	\begin{tabular}{cccccccccc}
		\toprule
		           &  \multicolumn{1}{p{1.5cm}}{\centering Material \\ density}
		           &  \multicolumn{1}{p{1.5cm}}{\centering Water  \\ density}
		            &\multicolumn{1}{p{1cm}}{ \centering Column \\ length}
		            &\multicolumn{1}{p{1cm}} { \centering  Tether \\ length}
		            & \multicolumn{1}{p{1cm}} { \centering  Ballast \\ position}
		             & \multicolumn{1}{p{1cm}} { \centering  Column \\ radius} 
		             &	\multicolumn{1}{p{1.5cm}} { \centering Column \\  thickness}
		             	&	\multicolumn{1}{p{1cm}} { \centering Ballast  \\ mass} 
		             	 &	\multicolumn{1}{p{1cm}} { \centering Mass  \\ coefficient}
		             	\\
		             
		\cmidrule(lr){2-10}
	               & $\rho$  [kg/m3] & $\rho_f$ [kg/m3]   & $L$ [m] & $L_S$ [m] & $L_b$ [m] & $r$ [m] & $t$ [m] & $m_b$ [kg] &$C_a$ [-] \\
		\cmidrule(lr){2-10}
		     Mean &  1180 & 1025 & 1 & 0.2 & 0.15 & 4.5e-2 & 3.5e-3 & 3 & 1     \\
	  \cmidrule(lr){2-10}            
			CoV 	&  \multicolumn{9}{c}{1e-4 ($\delta$-approximation with a small CoV)}
			 \\
		\bottomrule
	\end{tabular}
	}
	\label{tab:3}
\end{table}
In this section, we demonstrate the potential application of the proposed sensitivity framework for deterministic inputs, where a $\delta$-type distribution is assumed for approximation. In this example, the natural frequency sensitivity of a model floating column in a wave tank environment is analysed. This two-degrees-of-freedom floating system, as seen in Figure \ref{fig:1c}, is tethered to the bottom of the wave tank and a ballast mass is added to adjust the centre of gravity. This example is chosen here mainly to represent the commonly encountered natural frequency design problem. Furthermore, a closed-form sensitivity analysis for the natural frequencies, as given in Appendix \ref{appendix:b}, can be obtained straightforwardly as a benchmark. The example has been modelled using this hydrodynamic code \cite{yang_code_2022} where the added inertia effect of the water is considered. The partial derivatives of the mass and stiffness matrices, which are required for the analytical sensitivity results as seen in Appendix \ref{appendix:b}, are obtained using the symbolic differentiation module within Matlab. The code for this example study can be found in the address given in \nameref{sec:data}. 

The sensitivity results for the two natural frequencies, namely $\omega_1$ and $\omega_2$, are displayed in Figure \ref{fig:4} for the parameters listed in Table \ref{tab:3}. The `Deterministic sensitivity' shows the results from the analytical analysis, as given in Appendix \ref{appendix:b}, using the mean values of the input variables as the evaluation values. To use the proposed sensitivity framework, the input variables are assumed to be uncertain with a Gaussian distribution and the mean and CoV listed in Table \ref{tab:3}. The Fisher information matrix (FIM), based on Eq \ref{eq:12}, is formed for the two natural frequencies and the eigenvectors of the FIM with large eigenvalues then provide us with the sensitivity information. In order to compare with the analytical deterministic sensitivities, the FIM is calculated for each natural frequency separately, i.e., one-dimensional PDFs of each natural frequency rather than the joint PDF.  In this case, only the 1st eigenvectors of FIM are shown as only the 1st eigenvalue dominates. Very good agreement can be observed between the FIM results and the analytical sensitivities. In particular, the relative phase of the parameter $L_b$ is captured by the FIM despite its relatively low sensitivity. 
\section{Conclusions}
\label{sec:6}
A sensitivity matrix $\bf r$ is proposed as a new probabilistic sensitivity metric with respect to the input distribution parameters. The sensitivity of a wide range of commonly used uncertainty metrics, from moments of the uncertain output to the entropy of the entire distribution, can be formulated as an eigenvalue problem of the 2nd moment of the proposed sensitivity matrix. The resulting framework has a solid mathematical underpinning, is numerically efficient, and unifies the sensitivity analysis in a general but conceptually simple framework. And that is the main contribution of the present work.  

The proposed framework is derived analytically via a constrained maximization of the perturbation of the output uncertainties. On top of this mathematical foundation is the framework's conceptual simplicity, where its implementation only consists of two main steps, a Monte Carlo type sampling followed by solving an eigenvalue equation. Through the numerical examples, it is demonstrated that the sensitivity framework can be applied for the combined sensitivity analysis of multiple responses, even if the degree of correlation between the responses is unknown. This is in line with the findings for multiple correlated failure modes studied in \cite{yang_combined_2022}. In addition, using the failure sensitivity as an example, it is shown that a robust sensitivity analysis can be formed using entropy as a constraint and solved as a generalised eigenvalue problem. Furthermore, the Fisher information matrix, a special case of the proposed sensitivity metric, is shown to approximate deterministic sensitivities very closely using $\delta$-type distribution inputs. 

A key element of the framework is the sensitivity matrix $\bf r$. As its elements are the normalised partial derivatives of the expected utility of interest, the resulting eigenvectors based on the sensitivity matrix have direct sensitivity interpretations. Utilising the likelihood ratio/score function method, when a sampling approach is used, the expected utility and its derivatives can be obtained in a single simulation run. This allows efficient computation of the sensitivity matrix $\bf r$ and the corresponding $\Err$ matrix. 

The stochastic aspects of the output of interest have been considered implicitly in this study. For example, the stochastic responses along the beam in case 1 have been considered by setting the $g(\cdot)$ function as the r.m.s operation. Future work will consider a stochastic variant of the proposed framework, including sensitivities with respect to time-dependent inputs and treating the expected utility as stochastic by updating the expectation operation in Eq \ref{eq:5}. 
\section*{Acknowledgment}
This work has been funded by the Engineering and Physical Sciences Research Council through the award of a Programme Grant “Digital Twins for Improved Dynamic Design”, Grant No. EP/R006768. For the purpose of open access, the author has applied a Creative Commons Attribution (CC BY) licence to any Author Accepted Manuscript version arising. The author is grateful to Professor Robin Langley, University of Cambridge, for the support to publish this work.
\section*{Data availability statement}
\label{sec:data}
The datasets generated during and/or analysed during the current study are available in the GitHub repository: \\
\url{https://github.com/longitude-jyang/Probabilistic-sensitivity-framework}



 \bibliographystyle{unsrt}
  \bibliography{references}

\begin{appendices}
\section{The summary sensitivity index}
\label{appendix:a}

\renewcommand{\theequation}{\thesection.\arabic{equation}}
\setcounter{equation}{0}
The eigenvalue equation from Eq \ref{eq:9} can be rewritten as a matrix decomposition (as $\Err$ is symmetric):
\begin{equation} \label{eq:a1}
    \Err \mathbf =\bf Q \Lambda Q^{\T}
\end{equation}
where $\bf Q$ is the eigenvector matrix with $\mathbf{q}_i$ as its $i_{th}$ column and $\Lambda$ is the diagonal eigenvalue matrix. \\
Eq \ref{eq:a1} can be further expressed as a summation over the eigenvalues:
\begin{equation} \label{eq:a2}
   \begin{split}
         \Err  & = \sum_i \lambda_i \mathbf{q}_i \mathbf{q}_i^{\T} \\
                  & = \sum_i \lambda_i 
                                 \begin{bmatrix}
                                       q_{1i}^2  &  \dots & q_{1i}q_{ji}  & \dots \\
                                                          &  \ddots &     &                 \\
                                       \text{sym} &               & q_{ji}^2   &         \\
                                                      &                &              & \ddots
                                 \end{bmatrix}
    \end{split}
\end{equation}
Therefore, the sensitivity summary index in Eq \ref{eq:23}, where $s_j^2 = \sum_i \lambda_i q_{ji}^2$, is essentially the diagonal entries of the moment matrix.  
\section{Analytical derivation for natural frequency sensitivity}
\label{appendix:b}

\renewcommand{\theequation}{\thesection.\arabic{equation}}
\setcounter{equation}{0}
The natural frequency $\omega$ of a linear discrete vibration system can be found from the following eigenvalue equation:
\begin{equation} \label{eq:b1}
    \mathbf{K\bm{\upphi}} = \omega^2 \mathbf{M\bm{\upphi}} 
\end{equation}
where $\bf K$ and $ \bf M$ are the stiffness and mass matrices, $\bm {\upphi}$ is the eigenvector and $\omega^2$ is the corresponding eigenvalue. Without loss of generality, the eigenvectors are assumed to be mass normalised, i.e., $\bm{\upphi}^{\T}\mathbf{M} \bm{\upphi} = 1 $. Therefore,  the partial derivatives of the squared natural frequency with respect to the parameters are:
\begin{equation} \label{eq:b2}
\begin{split}
           \frac{\partial \omega^2}{\partial b_j} & = \frac{\partial}{\partial b_j} \left( \bm{\upphi}^{\T} \mathbf{K} \bm{\upphi} \right)    \\ 
           &  = \omega^2 \left( \bm{\upphi}_{b_j}^{\T} \mathbf {M} \bm{\upphi} +  \bm{\upphi}^{\T} \mathbf {M} \bm{\upphi}_{b_j} \right) + \bm{\upphi}^{\T} \mathbf {K}_{b_j} \bm{\upphi}     
\\
           &  = \omega^2 \left( - \bm{\upphi}^{\T} \mathbf {M}_{b_j} \bm{\upphi} \right) + \bm{\upphi}^{\T} \mathbf {K}_{b_j} \bm{\upphi}     \\
           &  = \bm{\upphi}^{\T} \left( \mathbf {K}_{b_j} - \omega^2 \mathbf{M}_{b_j} \right) \bm{\upphi} 
\end{split}
\end{equation}
where the subscript $b_j$ denotes the partial derivative with respect to the parameter $b_j$. In our case study with random input variables, the parameter $b_j$ is taken as the mean value of the $j_{th}$ variable. The 2nd to 3rd step of Eq \ref{eq:b2} makes use of the fact that $\frac{\partial}{\partial b_j} \left( \bm{\upphi}^{\T} \mathbf{M} \bm{\upphi} \right) = 0$. The partial derivatives of the natural frequency are then: 
\begin{equation} \label{eq:b3}
    \frac{\partial \omega}{\partial b_j} 
    = \frac{1}{2\omega} \bm{\upphi}^{\T} 
    \left[ \frac{\partial \mathbf{K}}{\partial b_j} - \omega^2 \frac{\partial \mathbf{M}}{\partial b_j} \right] \bm{\upphi}
\end{equation}
and the normalised sensitivity is: 
\begin{equation} \label{eq:b4}
   r_{\omega}^j= \frac{\partial \omega}{\partial b_j} \frac{b_j}{\omega}
\end{equation}

\end{appendices}

%
%
%
%
\end{document}